\documentclass[copyright,creativecommons,noderivs,noncommercial]{eptcs}
\providecommand{\event}{FBTC 2010} 

\usepackage{breakurl}             
\usepackage{verbatim}
\usepackage{graphicx}
\usepackage{amsmath, amssymb}
\usepackage{txfonts}
\usepackage{empheq}
\usepackage{setspace}

\usepackage{pepa}
\usepackage{subfigure}

\newcommand{\Real}{{\mathbb R}}

\newcommand{\reactant}{\mbox{$\downarrow$}}
\newcommand{\product}{\mbox{$\uparrow$}}

\newcommand{\modifier}{\odot}
\newcommand{\activator}{\oplus}
\newcommand{\inhibitor}{\ominus}
\newtheorem{definition}{\textbf{Definition}}

\newcommand{\syncstar}[1]{\raisebox{-1.0ex}{$\;\stackrel{\S}{\scriptstyle #1}\,$}}

\title{Investigating modularity in the analysis of process algebra
models of biochemical systems}
\author{
Federica Ciocchetta${}^{1}$ \quad\quad
Maria Luisa Guerriero${}^{2}$ \quad\quad
Jane Hillston${}^{2,3}$ \\
\institute{${}^{1}$ The Microsoft Research - University of Trento Centre for
Computational and Systems Biology, Trento, Italy\\
${}^{2}$ Centre for Systems Biology at Edinburgh, University of
Edinburgh, UK\\
${}^{3}$ School of Informatics, University of Edinburgh, UK}}

\def\titlerunning{Investigating modularity in the analysis of process
algebra models of biochemical systems}
\def\authorrunning{F. Ciocchetta, M.L. Guerriero \& J. Hillston}
\begin{document}
\maketitle

\begin{abstract}

Compositionality is a key feature of process algebras which is often
cited as one of their advantages as a modelling technique.  It is
certainly true that in biochemical systems, as in many other
systems, model construction is made easier in a formalism which
allows the problem to be tackled compositionally.  In this paper we
consider the extent to which the compositional structure which is
inherent in process algebra models of biochemical systems can be
exploited during model solution.  In essence this means using the
compositional structure to guide decomposed solution and analysis.

Unfortunately the dynamic behaviour of biochemical systems exhibits
strong interdependencies between the components of the model making
decomposed solution a difficult task.  Nevertheless we believe that if
such decomposition based on process algebras could be established it
would demonstrate substantial benefits for systems biology modelling.
In this paper we present our preliminary investigations based on a
case study of the pheromone pathway in yeast, modelling in the
stochastic process algebra Bio-PEPA.




\end{abstract}

\section{Introduction}
\label{sec:intro}
Biochemical systems are generally large and complex and, therefore,
studying a biochemical system monolithically can be difficult, both
in terms of the definition of the model and in terms of its
analysis. Over the last decade process algebras have been proposed
as suitable formalisms for constructing models of biochemical
systems~\cite{priamiEtAl01,regevEtAl04,priami-quaglia05b,ciocchetta-hillston09},
with their compositional structure being claimed as one of their
main advantages.

In many cases the process algebra model is being used as an
intermediate language which gives rise to a mathematical
representation of the dynamics of the system, on which analysis of
the behaviour can be conducted.  In many cases this mathematical
representation is a stochastic simulation based on an implicit
continuous time Markov chain (CTMC), but in others it may be an
explicit CTMC, a set of ordinary differential equations (ODEs) or
some combination of these.  However, regardless of the mathematical
formalism chosen, the size and complexity of the biochemical
processes can give rise to problems of tractability in the
underlying mathematical model.  This problem is particularly acute
in the case of explicit CTMC models such as those used in
probabilistic model-checking.  Thus it is attractive to consider the
extent to which decomposed model analysis can be used.  This
approach considers the model to be comprised of a number of modules
which may be analysed in isolation, their results being combined to
give results for the complete systems.  Furthermore when the model
has been described in a compositional formalism we would hope that
the compositional structure may be exploited in the identification
of suitable modules. In this work we refer to process algebras
because they are generally equipped with a formal compositional
definition, which we can rely on for our modular analysis. It is
worth noting, however, that the same approach could potentially be
applied to any other language equipped with a notion of modules.

Deriving global dynamic behaviour from the study of isolated
components is difficult but the rich potential benefits mean that it
has been studied in a wide variety of contexts.  For example, notions
of modularity and model decomposition have been widely applied in the
context of software engineering, network theory and, more recently,
systems biology
(e.g.~\cite{saezrodriguez-kremling-gilles05,Rodriguez04,conzelmannEtAl04,Bruggeman08,Wolf03,Hartwell99}).
Moreover, model decomposition techniques, such as \textit{product form
approaches} and \textit{time scale decomposition}, have been defined
and applied in the context of CTMC-based analyses and process
algebras~\cite{Hillston01,Hillston95}.

However, the full strength of compositionality of process algebra
has yet to be used in the context of quantitative analysis of
biochemical models, although there have been some attempts to
exploit the compositional structure in qualitative analysis and via
congruence relations (for example,
see~\cite{barbutiEtAl08,pintoEtAl07}). In this paper we present a
case study of decomposed model analysis. The system we consider is
the yeast pheromone system, previously studied by Kofahl and
Klipp~\cite{Kofahl04}.  In that paper the authors present an ODE
model and an informal decomposition into modules. Here we give a
formal model of the system in the stochastic process algebra
Bio-PEPA~\cite{ciocchetta-hillston07,ciocchetta-hillston09} and
demonstrate how the compositional structure of the process algebra
supports a rigorous definition of modules.  Moreover the flexibility
of the Bio-PEPA framework to generate a number of different
underlying mathematical models allows different analysis techniques
to be used in tandem to parameterise and analyse the submodels
corresponding to the modules.  We choose as our main focus the use
of probabilistic model-checking to verify properties of the pathway,
partly because this style of analysis is a feature of Bio-PEPA but
also because this requires an explicit state space CTMC, so the
problem of state space explosion is particularly acute. 

As suggested above, we adopt a high-level language, such as
Bio-PEPA, instead of considering directly the CTMC model, in order
to take advantage of the multiple analysis techniques it supports,
since these can be employed in a complementary fashion during model
analysis. Specifically, we first use stochastic simulation in order
to validate our model against the results in the literature and
derive some information necessary for the definition of the
associated explicit CTMC submodels. Subsequently, we study further
properties of the submodels using model-checking and investigate
which properties of the submodels can safely be interpreted in the
context of the complete model. It is worth noting that the
decomposition of the Bio-PEPA model into modules does not require us
to build the full CTMC model, which indeed can be very large and
difficult to represent and study explicitly.

The rest of the paper is organised as follows. First, we present a
discussion of model decomposition and an overview of our approach
(Section~\ref{sec:modularity}).  In the following section
(Section~\ref{sec:Bio-PEPA}) we give a brief introduction to the
Bio-PEPA language. Our approach is applied to the yeast pheromone
pathway in the following Sections~\ref{sec:pheromone}
and~\ref{sec:BPmodel}. Some conclusions and future work are reported
in Section~\ref{sec:conclusions}.

\section{Decomposing models}
\label{sec:modularity}

As mentioned earlier, a process algebra description of a biochemical
pathway can be regarded as an intermediate representation sitting
between the biological knowledge and the mathematical models on which
analysis is conducted.  Consequently the process algebra description
plays two roles.  Firstly, it is a repository of the current
biological knowledge about the system under study.  Secondly, it is a
specification of the mathematical model which is to be constructed to
analyse that system.

When we wish to consider the decomposition of the process algebra
model it follows that we may view the problem from the two different
perspectives: the biological and the mathematical.  In the biological
perspective the existence of functional units within biochemical
systems is well known.  Perhaps the most intuitive examples of this
functional separation are signalling pathways, which can be considered
as units responsible for conveying specific messages within larger
systems.  Nevertheless, these modules often affect each other, either
positively or negatively, through crosstalk.

Some attempts at defining modules with biochemical systems via
biological function have appeared in the literature recently, for
instance~\cite{Wolf03,Hartwell99,Bruggeman08,saezrodriguez-kremling-gilles05,Rodriguez04,conzelmannEtAl04}.
In~\cite{saezrodriguez-kremling-gilles05,Rodriguez04,conzelmannEtAl04}
Saez-Rodriguez \textit{et al.}  decompose signalling pathways into
modules, study them in isolation using ODE-based analysis and derive
global properties of the system from the local behaviours.  Their
model decomposition approach is based on the structure of
biochemical networks, in particular on the definition of
``independent units''. The independence of modules is expressed in
terms of \emph{absence of retroactivity} in the interaction between
modules.
A connection between two modules is retroactivity-free if it does
not involve direct interactions in both directions. If there are
indirect interactions, such as feedbacks, the connection between
modules exhibits \emph{weak retroactivity}. The input/output
behaviour of a retroactivity-free module does not depend on what it
is connected to; therefore, if retroactivity-free functional
subunits can be identified within a system, we obtain a
modularisation in which the independence of modules is inherent to
the system's functional subunits. The notion of modules defined  by
Bruggeman \textit{et al.} in~\cite{Bruggeman08} is slightly more
specific: the authors distinguish between \textit{levels} and
\textit{modules}. Subunits which only have \textit{regulatory}
interactions (a species in a subunit acts as an enzyme/inhibitor of a
reaction within another subunit, i.e.~interactions do not involve
mass-transfer between subunits) are called levels; instead, subunits
whose interactions involve some mass-transfer (a species in a subunit
acts as a product/reactant of a reaction within another subunit) are
called modules.

In contrast, from the mathematical perspective the decomposition of
the system is generally based on notions of interaction and
independence at the state level.  This is quite different because
interactions between states (often based on the count of molecules
of each species) are orthogonal to the interactions between species.
In this context various decompositional techniques have been
proposed in the context of CTMC models and process algebras in order
to aid in the solution of large Markov processes~\cite{Hillston01}.
For example \textit{product form techniques} rely on only restricted
forms of state transitions but do mean that important probability
measures relating to the whole system can be retrieved exactly from
the analysis of modules in isolation.  Other decomposition
techniques are generally approximate and may involve solution of an
aggregated model recording the interactions between modules in
addition to the analysis of the isolated modules.  One example would
be \textit{time scale decomposition} of the CTMC model: modules are
formed by grouping states in the CTMC which can be reached quickly
relatively to the transition rates to states in other
modules~\cite{busch-sandmann-wolf06}.

Of course, ideally we would like these two views of modules arising
from a stochastic process algebra model of a biochemical pathway to
coincide.  One of the objectives of this paper is to investigate the
extent to which that occurs by taking a biologically inspired
decomposition and studying how well it performs as a mathematical
decomposition.

Which ever approach to decomposition we are taking, in a
compositional modelling language the task is more straightforward if
the boundaries between the modules required for decomposed solution
coincide with the boundaries between components used to construct the
model.  As we will see this is indeed the case of Bio-PEPA models.
In our study each module is considered as an individual smaller
network: its behaviour in isolation is analysed, and its local
properties identified. When looking at one module in
isolation, the behaviour of the other modules can be
regarded as part of the external environment. Therefore, the
effect of changes in the other modules on the module under study
must be investigated. This first analysis step allows us to
understand in depth the local behaviour of each part of the pathway.

It is a challenge to identify local properties that
can help in understanding the behaviour of the full system. The main
issue is to understand \emph{which} properties can be considered and
\emph{when} they can be applied. Structural properties are in
general suitable because the structure of the system is not altered
by the model composition; dynamic properties, instead, might not be
appropriate, because the system's dynamics is not guaranteed to be
preserved by the composition, especially if feedback mechanisms
involving different modules are present.
In~\cite{saezrodriguez-kremling-gilles05,Rodriguez04,conzelmannEtAl04},
for instance, the authors focus on some properties used in network
theory that can be useful also in the context of signalling
pathways, such as \emph{signal amplitude}, \emph{signalling time},
and \emph{signal duration}.  From the mathematical perspective the
focus is generally on the transient and steady state probability
distributions over states.

\begin{figure}[t]
\centering
\includegraphics[width=0.74\textwidth]{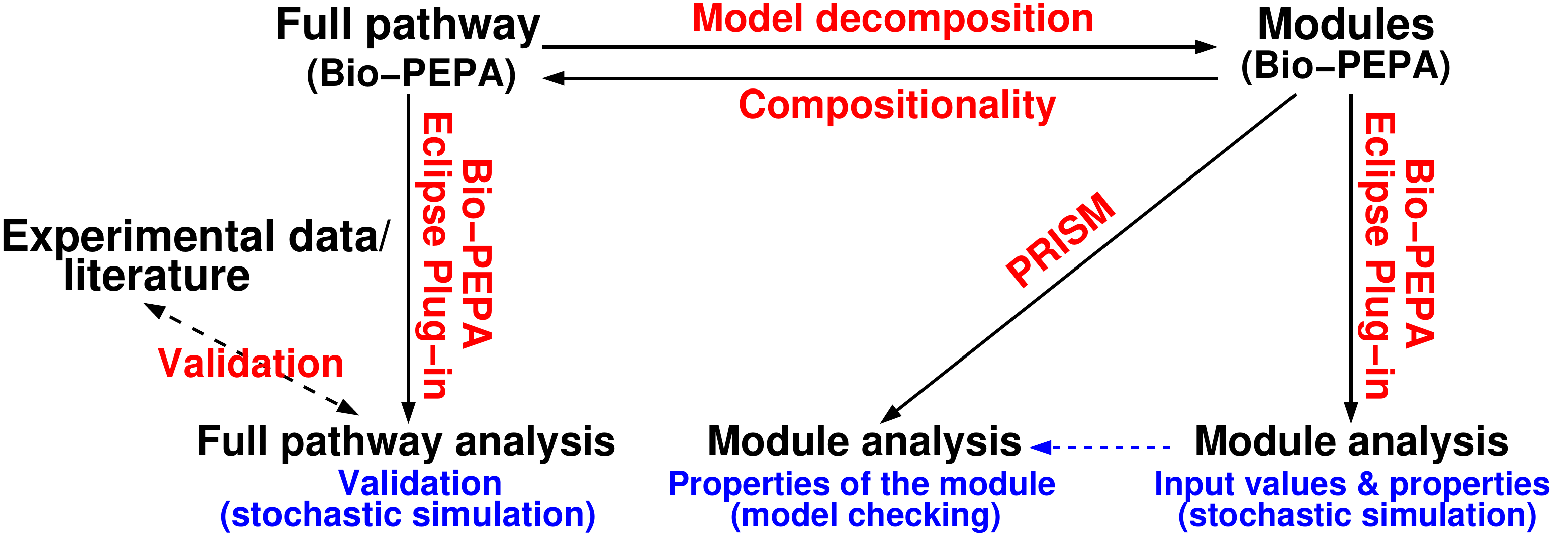}
\caption{\label{fig:approach} Approach considered in this paper.}
\end{figure}

As already stated, in our investigation we exploit the compositional
and modular structure of Bio-PEPA models and the variety of analysis
techniques supported by the language. The schema of the approach
considered is reported in Figure~\ref{fig:approach}.
We define a Bio-PEPA model for the system and we validate it, as a
single entity, against existing data from experiments or literature.
The model is then decomposed into modules for detailed analysis,
using stochastic simulation and model-checking. The analysis of
modules via stochastic simulation is helpful to study some
preliminary properties and derive the
information needed for CTMC analysis: 
for this purpose we use the Gibson-Bruck stochastic simulation
algorithm~\cite{gibson-bruck00} available in the Bio-PEPA Eclipse
Plug-in~\cite{biopepa_site}. The PRISM
model-checker~\cite{prism_site} is then used to verify further local
properties on the individual modules that can be difficult or
impossible to observe directly from stochastic simulation. Moreover,
differently from simulation, probabilistic model-checking is an
exhaustive approach and, thus, it allows us to simultaneously
consider all the possible stochastic behaviours of the modules
rather than a finite subset of them. Note that the considered local
properties are preserved in the full system, and their verification
on the model of the full system is not feasible due to its size.

In order to correctly parameterise the modules in isolation and to
interpret the results that we obtain from them, it can be useful to
classify the relationship between the modules in terms of their
local species and the ones which are on the interface between
different modules and act as input/output species. Specifically,
when considering a module in isolation, its species can be either
(1) \textit{local} (i.e.~not present elsewhere) or (2) involved in
reactions occurring in different modules; in the second case, they
can act in other modules as either (2a) \textit{external regulators}
(i.e.~activators, inhibitors or modifiers) or (2b) \textit{external
reagents} (i.e.~products/reactants). Case (2a) corresponds to the
notion of \emph{level} proposed in~\cite{Bruggeman08}.


When analysing a single module in isolation, this must be
parameterised to emulate the effect of the external environment
(i.e.~the other modules). The hardest case to take into account is
when there are species of type (2b), since this means that the
amount of the species is changed by the external environment rather
than being independently regulated within the considered module. For
this type of non-local species, it is crucial to choose appropriate
initial values, and additional creation or degradation reactions for
such species might be required.


Our objective here is to explore these issues empirically.
Based on our experience we believe that an automated procedure could
be implemented based on parameter estimation software. This would
support finding optimal values for the input/output species and
their creation/degradation reactions by comparing the results of the
modules in isolation with the results of the whole system and/or
experimental data.

\section{Bio-PEPA}
\label{sec:Bio-PEPA} In this section we give a short description of
Bio-PEPA, a language that has recently been developed for the
modelling and analysis of biological systems.  The interested reader
is referred to~\cite{ciocchetta-hillston07,ciocchetta-hillston09}
for more details.
The main components of a Bio-PEPA system are the \emph{species
components}, describing the behaviour of each species, and the
\emph{model component}, describing the interactions between the
various species.
The syntax of Bio-PEPA components is:
$$
S ::= (\alpha, \kappa) \mbox{ \texttt{op} } S \mid S + S \mid C
\quad  \mbox{with }\texttt{op} = \reactant \mid \product \mid
\activator \mid \inhibitor \mid \modifier \quad \quad \quad P::=P
\sync{\mathcal{I}}  P \mid S(x)
$$
\noindent where $S$ is the \emph{species component} and $P$ is the
\emph{model component}. In the prefix term $(\alpha,\kappa) \mbox{
\texttt{op} } S$, $\kappa$ is the \emph{stoichiometry coefficient}
of species $S$ in reaction $\alpha$, and the \emph{prefix
combinator} ``$\texttt{op}$'' represents the role of $S$ in the
reaction. Specifically, $\reactant$ indicates a \emph{reactant},
$\product$ a \emph{product}, $\activator$ an \emph{activator},
$\inhibitor$ an \emph{inhibitor}, and $\modifier$ a generic
\emph{modifier}. We use the notation ``$\alpha \mbox{ \texttt{op} }
S$'' as an abbreviation for ``$(\alpha, \kappa) \mbox{ \texttt{op} }
S$'' when $\kappa=1$. The operator ``$+$'' expresses the choice
between possible actions, and the constant $C$ is defined by an
equation $C \rmdef S$. The process $P \sync{\mathcal{I}} Q$ denotes
synchronisation between components $P$ and $Q$, the set
$\mathcal{I}$ determines those activities on which the operands are
forced to synchronise, with $\syncstar{*}$ denoting a
synchronisation on all common action types.
In  $S(x)$  the parameter $x \in \Real$
represents the initial value for species $S$.

Biological locations (representing both compartments and membranes)
can be defined, and the notation $C @ L$ indicates that species $C$
is in location $L$. For additional details of the definition of
locations see~\cite{Ciocchetta-Guerriero09}.
The formal definition of a Bio-PEPA system is the following.
\begin{definition}
\label{def:biopepa} A Bio-PEPA system $\mathcal{P}$ is a 6-tuple
$\langle \mathcal{L},\mathcal{N},\mathcal{K}, \mathcal{F}_R,Comp,P
\rangle$, where: $\mathcal{L}$ is the set of locations,
$\mathcal{N}$ is an optional set of information for the species,
$\mathcal{K}$ is the set of parameters, $\mathcal{F}_R$ is the set
of functional rates, $Comp$ is the set of species components,
and $P$ is the model component.
\end{definition}

The language is given a formal, discrete state, small step semantics
based on operational rules ---
see~\cite{ciocchetta-hillston07,ciocchetta-hillston09} for
details. Moreover, mappings have been defined from Bio-PEPA to a
variety of underlying mathematical models and analysis techniques
--- stochastic simulation based on CTMC, ordinary differential
equations (ODEs) and probabilistic model-checking.   These mappings
and the subsequent analyses are supported by various software
tools~\cite{biopepa_site,dizzy_site,prism_site}. 
In the case of
numerical analysis of CTMC and probabilistic model-checking using
PRISM~\cite{KNP09a,prism_site}, species are abstracted in terms of
levels, each level representing an interval $h$ of values
(representing number of molecules, for instance); the set
$\mathcal{N}$ specifies the information needed for the derivation of
the CTMC with levels, such as the lower and upper bounds on
molecular amounts, and the step sizes $h$. In PRISM quantitative
properties of the system are expressed using the temporal logic
\textit{CSL} (Continuous Stochastic Logic)~\cite{ASSB96} and
rewards.

\section{The yeast pheromone pathway}
\label{sec:pheromone} 
\begin{figure}[t]
\centering
\includegraphics[width=0.6\textwidth]{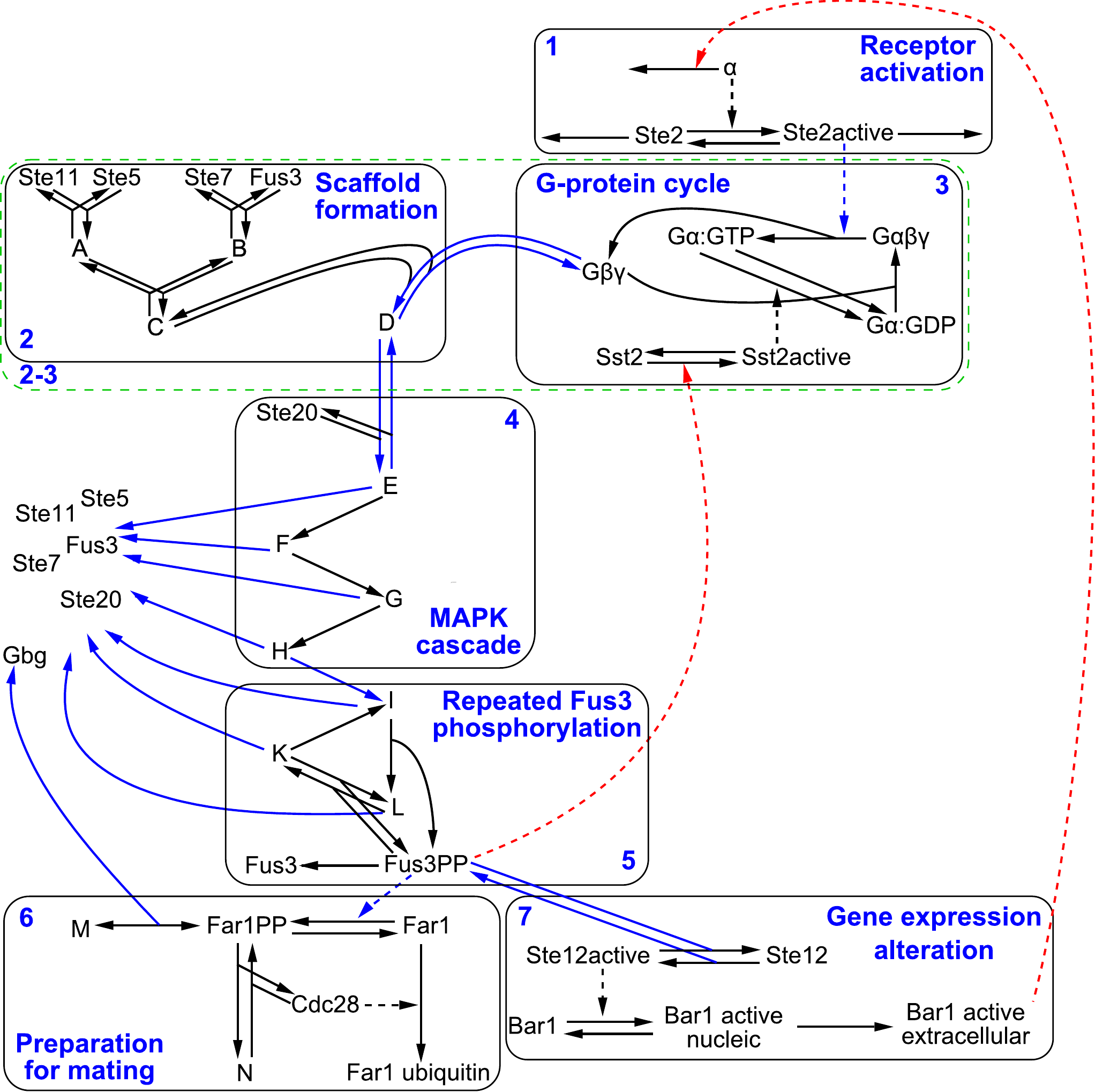}
\caption{\label{fig:model_schema}Schema and module decomposition of
the yeast pheromone pathway as described in~\cite{Kofahl04}. 
Black arrows represent interactions which are local to a module, red
arrows show feedback loops, blue arrows represent reactions where the
outputs of a module are reagents or regulators in another module.
Dashed lines of any colour describe regulatory interactions between
species whereas solid lines represent interactions involving reagents.
The black boxes are the modules defined in~\cite{Kofahl04}, and the
green dashed box represents the module obtained by merging modules 2
and 3.}
\end{figure}

In order to illustrate our approach, we
consider the \textit{yeast pheromone pathway} in haploid yeast cells
after stimulation by $\alpha$-factor~\cite{Kofahl04}.
A schema of the pathway together with the definition of modules presented in~\cite{Kofahl04} and discussed below  is reported in Figure~\ref{fig:model_schema}. The
receptor Ste2 is activated by the pheromone $\alpha$-factor and in
turn activates a heterotrimeric G-protein which transmits the signal
from the cell surface receptor to intracellular effectors. In
particular, the receptor interacts with some subunits of $G$. This
leads to a series of conformational changes, allowing, among others,
the release of $G\beta\gamma$. $G\beta\gamma$ binds to and activates
a scaffold protein-bound mitogen-activated protein kinase (MAP
kinase complex, $C$ in the figure), resulting from the mating
pheromone subpathway (top left part in the figure). The activation
of the MAPK cascade (from complex $D$ to $G$) follows, together with
further intermediate reactions that lead to the activation of
nucleic proteins that control transcription and progression in the
cell cycle. Specifically, the protein Fus3 is activated by double
phosphorylation at the end of the MAPK cascade. Activated Fus3
phosphorylates and regulates various proteins, such as Ste12 and
Sst2. The activation of Ste12 leads to the activation of the protein
Bar1, which is translocated in the extracellular space and inhibits
the activity of the $\alpha$-factor by degradation.

We have two \emph{feedback loops} in the model to take into account:
\begin{enumerate}
\item the activated Bar1 in the extracellular space
down-regulates the activation of the receptor by enhancing the
degradation of the $\alpha$-factor. Bar1 is activated in the last
part of the pathway and its quantity is affected by the upstream
pathway.
\item Fus3 is activated by double phosphorylation in the last part
of the pathway and it is used to activate Sst2; this reduces the
quantity of G-protein available for the formation of signalling
complexes.
\end{enumerate}

We consider the deterministic model of this pathway defined
in~\cite{Kofahl04}. This involves four locations
(\emph{extracellular space, cytoplasm, cell membrane} and
\emph{nucleus}), 35 species and 45 reactions. Most reactions follow
the mass-action kinetic law, and a few follow Hill kinetics.
%
%
%
The following informal model decomposition of the pathway, driven by
the biological meaning of its subparts, is described
in~\cite{Kofahl04}  and reported schematically in
Figure~\ref{fig:model_schema} (the modules are represented by the
black boxes). 
\begin{description}
  \item [Module 1: Receptor activation.] The pheromone $\alpha$ factor
activates the receptor Ste2.
 \item [Module 2: Scaffold formation.] Formation of the
scaffold complex for the following MAPK cascade. 
 \item [Module 3: G-protein cycle.] This describes the activity of the
G-protein.
 \item [Module 4: MAPK cascade.]  This module is composed of three
phosphorylation steps, 
and in each of them one species phosphorylates another downstream
one. The final result is an active complex containing double
phosphorylated Fus3.
\item [Module 5: Repeated Fus3 phosphorylation.] After the MAPK cascade
there are a few further reactions that lead to the release of double
phosphorylated Fus3.
\item [Module 6: Preparation for mating.] This module describes one of
the possible activities influenced by the pheromone pathway.
\item [Module 7: Gene expression alteration.] Phosphorylated Fus3 activates Ste12
that in turn activates Bar1.
\end{description}

According to the described
modularisation, modules 2 and 3 are strongly interconnected (via
reversible reactions) and, hence, not independent. We consider a
modified modularisation, in which modules 2 and 3 are merged in a
new module (called \textit{module 2-3}, the dashed green box in
Figure~\ref{fig:model_schema}) while all other modules are
unchanged.


\section{Modelling and analysis of the pathway in Bio-PEPA}
\label{sec:BPmodel}

In this section we show how the pheromone pathway~\cite{Kofahl04}
can be modelled exploiting the compositionality offered by the
Bio-PEPA language, and we apply decomposed model analysis to the
pheromone model. We consider the modules as described in the
previous section and we analyse each of them individually. The first
issue we need to address in order to analyse modules in isolation is
to identify valid initial conditions and parameter values; we show
the correctness of the modularisation by comparing the simulation
results obtained for the individual models with the ones obtained
for the complete model. We also report some results obtained using
the PRISM model-checker for the verification of local properties
which are valid also globally.

Due to space constraints, we show here the definition and the analysis
of two of the modules, i.e.~module 1 and module 7, which represent the
``start'' and the ``end'' of the signalling pathway,
respectively. These modules are interesting because they are
interconnected by one of the two feedback loops present in the
system. Similar considerations can be used for the other modules to
obtain full understanding of the model. The full model is available
from~\cite{biopepa_site} and can be directly imported into the
Bio-PEPA Eclipse Plug-in. Note that in order to adhere to the syntax
of the formal definition of Bio-PEPA as presented in
Section~\ref{sec:Bio-PEPA}, the following description is slightly
different from the concrete syntax accepted by the tool. The reader is
referred to~\cite{duguid09} for the details of the syntax and usage of
the tool.

Each biochemical species of the pathway is represented by a species
component, and each biochemical reaction is described by an action
type and it is associated with a functional rate describing its
kinetic law. Auxiliary information about species, such as the step
size and the maximum molecular count, are defined in the set
$\mathcal{N}$. Locations, kinetic parameters, functions and
observables can be also defined. The quantitative parameters
(kinetic constants and initial values) are the ones used
in~\cite{Kofahl04}, and are not reported here.

The four different locations relevant to the pathway are abstracted
using following set of locations:
$$\mathcal{L} = [  \; extra : 1 \, \mathrm{\mu l}, \mathbf{C} ;  \; \ cyto : 1 \, \mathrm{\mu l} , \mathbf{C} ;  \; \ nucl  : 1 \, \mathrm{\mu l},  \mathbf{C};  \;
mem : 1 \, \mathrm{\mu l}, \mathbf{M}  \; ] \enspace .$$

The model component contains all the species components of the
systems with their initial values. Since the model component can be
defined in a compositional way, we can represent the full pathway as
the composition of the different modules. For each module we
consider two subsets of species, namely the local species and the
input/output ones. In the next sections we show how each of them is
defined for the two considered modules.


\subsection{Module 1: activation of the receptor}
\label{sec:module1}

\begin{figure}[hbtp]
\centering
\includegraphics[width=0.25\textwidth]{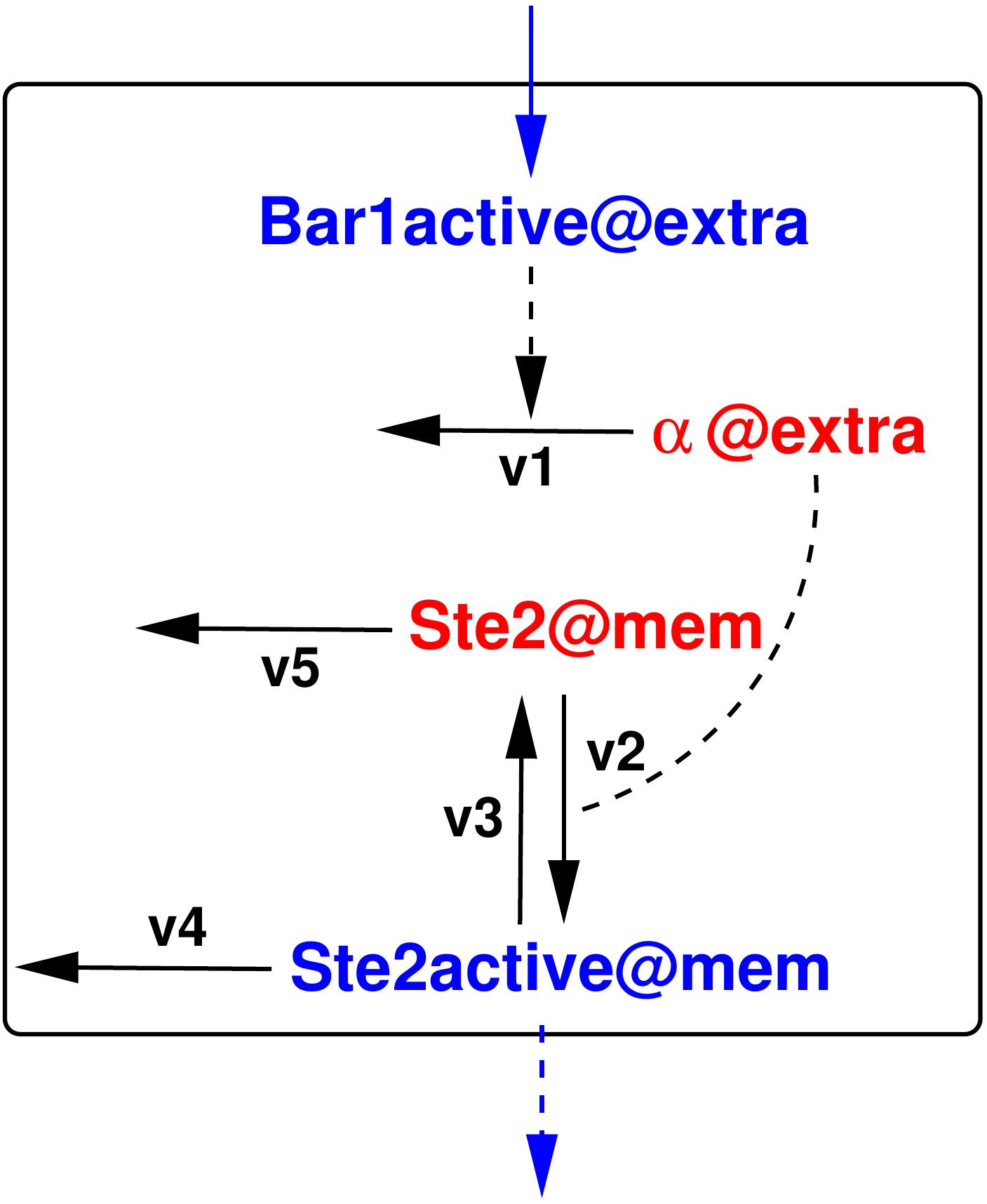}
\caption{\label{fig:module1} Description of module 1. The species in
red are the local species, while the species in blue the external
ones. Black arrows are generic reactions in the module whereas the
dashed lines represent activation/inhibition. Blue lines are
reactions to/from other modules.}
\end{figure}

The first module we consider is the one describing the receptor
activation (Figure~\ref{fig:module1}).

The only input species from outside the module is the active
extracellular Bar1 ($Bar1active@extra$), generated in module 7, which is an external
reagent, whereas the active Ste2 receptor ($Ste2active@mem$) is an
external regulator in module 2-3. The local species, which are not
influenced by other modules, are the $\alpha$-factor ligand
($\alpha@extra$) and the inactive Ste2 receptor ($Ste2@mem$).

The species definitions for the species of the module are the
following
\begin{center}
\begin{tabular}{lcl}
$\alpha@extra$ & $\rmdef$ & $v_1 \,\reactant \,\alpha@extra \; +  \; v_2 \,\activator \,\alpha@extra$ \\
$Ste2@mem$ & $\rmdef$ & $v_2 \,\reactant \,Ste2@mem \; + \; v_5 \,\reactant\, Ste2@mem \; + \; v_3 \, \product\, Ste2@mem$ \\
$Ste2active@mem$ & $\rmdef$ & $v_2 \,\product \,Ste2active@mem \; + \; v_4 \, \reactant\, Ste2active@mem \; +  \; v_3 \, \reactant \,Ste2active@mem$ \\
$Bar1active@extra$ & $\rmdef$ & $v_1 \, \activator \,Bar1active@extra \; +  \; v_{38} \, \product \,Bar1active@extra$
\end{tabular}
\end{center}
\noindent and the functional rates for these reactions are:
$$f_{v_1} = fMA (k_1); \; \; \; f_{v_2} = fMA (k_2); \; \; \; f_{v_3} = fMA
(k_3); \; \; \; f_{v_4} = fMA (k_4); \; \; \; f_{v_5} = fMA (k_5);
\; \; \; f_{v_{38}} = fMA (k_{38})$$ \noindent where $fMA(r)$ stands
for mass-action law with kinetic constant $r$, $v_1$ represents the
degradation of the $\alpha$-factor, $v_2$ the activation of the
receptor, $v_3$ its deactivation, $v_4$ the degradation of the
active receptor, $v_5$ the degradation of the inactive receptor, and
$v_{38}$ the transport of $Bar1active$ from the nucleus to the
extracellular space. The parameters $k_1, \dots, k_5$ and $k_{38}$
are defined as in~\cite{Kofahl04}.

The model component describing the full module 1 is the following.
$$\begin{array}{lcl}
Module1\_local \!\!\! & \!\!\! \rmdef \!\!\! & \!\!\! alpha@extra[init\_alpha\_extra] \syncstar{*} Ste2@mem[init\_Ste2\_mem] \\
Module1\_I/O   \!\!\! & \!\!\! \rmdef \!\!\! & \!\!\! Bar1active@extra[init\_Bar1active\_extra] \syncstar{*} Ste2active@mem[init\_Ste2active\_mem]
\end{array}$$

The initial values are the same as in~\cite{Kofahl04} for all the
species except $Bar1active@extra$ which, being an external reagent,
deserves further discussion.

Bar1 is inactive in the full pathway, and it is activated and
transported into the extracellular environment after the activation
of Fus3 by double phosphorylation (in module 7). Reaction $v_{38}$,
represents the transport of $Bar1active$ from the nucleus to the
extracellular space, which is in the interface between module 7 and
module 1. The rate of this reaction depends on the amount of
$Bar1active@nucl$, which is not part of this module. Therefore, when
considering module 1 in isolation, we need to find an appropriate
way to parameterise this reaction, and to assign an initial value to
$Bar1active@extra$: these values must be chosen in order to make the
behaviour of the species in the module the same as their behaviour
in the model of the full pathway.

\begin{figure}[hbtp]
\centering \subfigure[Full
pathway]{\includegraphics[width=0.32\textwidth]{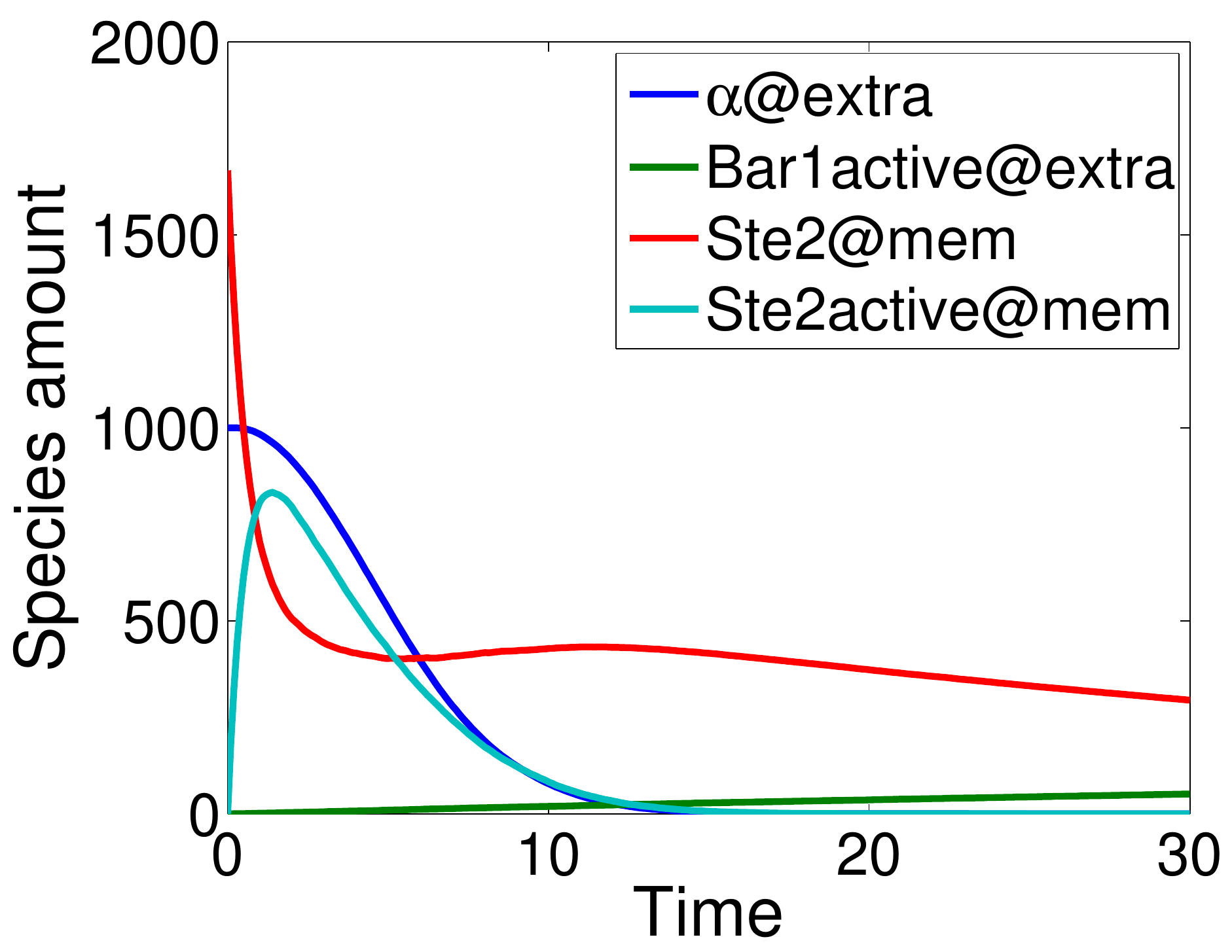}\label{fig:module1_ssa:full}}
\subfigure[$init\_Bar1active@extra=25$]{\includegraphics[width=0.32\textwidth]{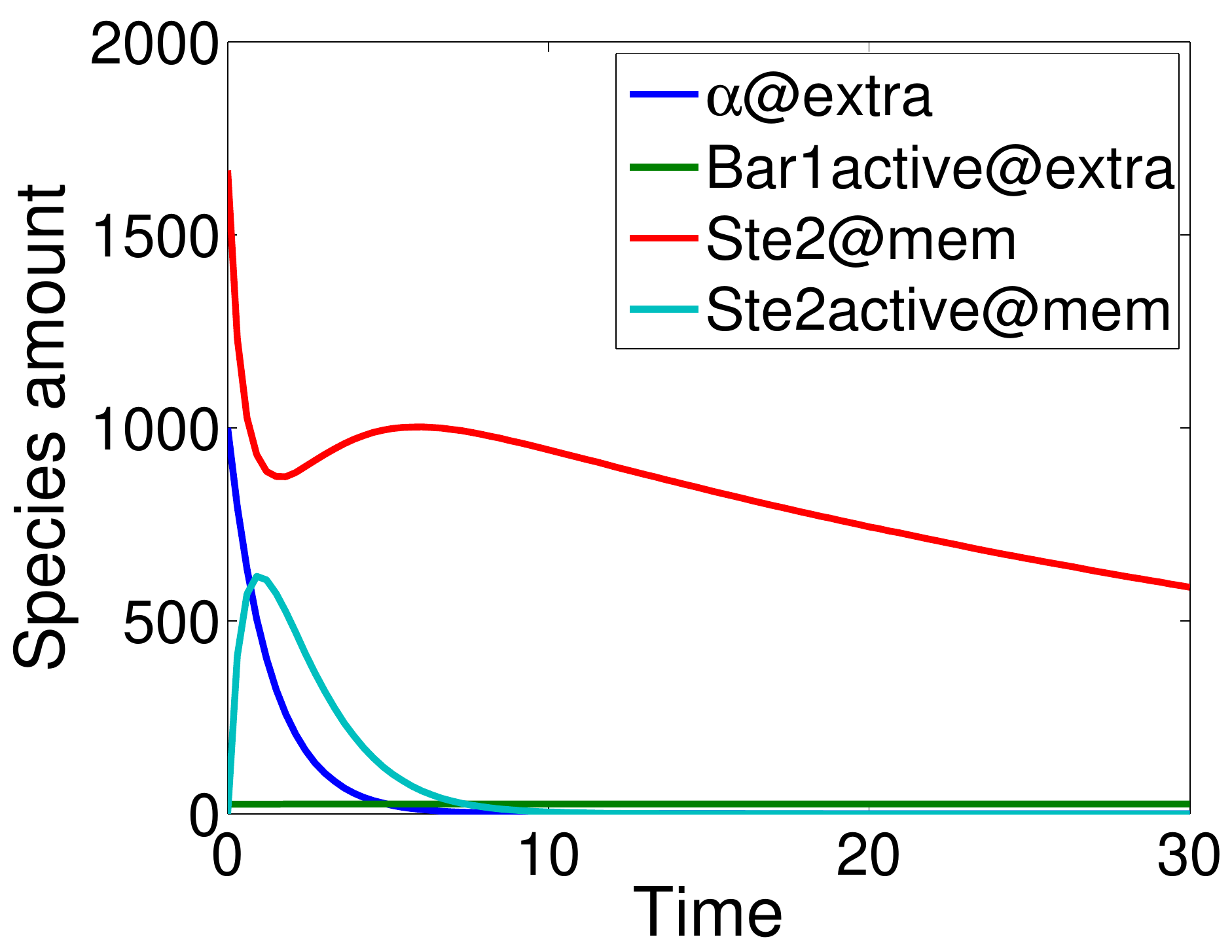}\label{fig:module1_ssa:init_25}}
\subfigure[Creation rate
$k_0=1.66$]{\includegraphics[width=0.32\textwidth]{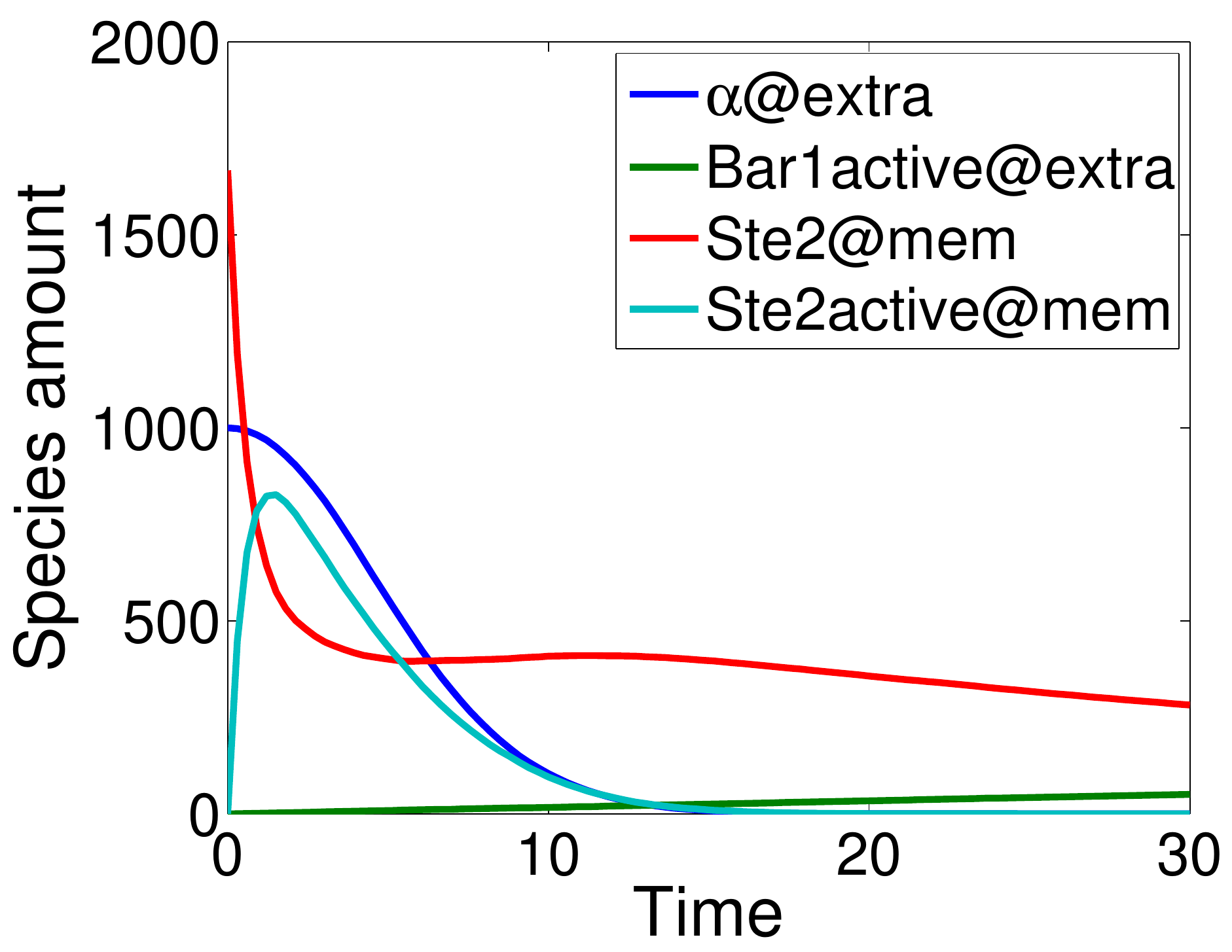}\label{fig:module1_ssa:creation_1.66}}
\caption{\label{fig:module1_ssa} Module 1: Time-series evolution
(Gibson-Bruck method, average of 100 simulation runs) of the
involved species for different assumptions about $Bar1active@extra$:
(a) full pathway, (b) module 1, with initial amount of
$Bar1active@extra=25$, and (c) module 1, with $Bar1active@extra$
initially absent and created with rate $k_0 = 1.66$.}
\end{figure}

If we interpret the species $Bar1active@extra$ as an \textit{input
signal}, we can study the effect of varying its value on the
behaviour of the module. We consider different possibilities for
$Bar1active@extra$, two of which are reported in
Figure~\ref{fig:module1_ssa}. This figure shows the average
stochastic simulation results (over 100 runs) for the species
involved in the module when the full pathway is considered
(Figure~\ref{fig:module1_ssa:full}), and for the module in isolation
(Figures~\ref{fig:module1_ssa:init_25}--\ref{fig:module1_ssa:creation_1.66}).
When the initial amount of $Bar1active@extra$ is equal to $25$ (half
the maximum value of the species obtained from simulation of the
whole system in the given interval of time,
Figure~\ref{fig:module1_ssa:init_25}), $Ste2active@mem$ has the
expected qualitative behaviour, but the peak is much lower and
$Ste2@mem$ does not decrease as expected. On the other hand, under
the assumption that $Bar1active@extra$ is initially absent and it is
created with a constant rate equal to $1.66$
(Figure~\ref{fig:module1_ssa:creation_1.66}), the results for the
module are in perfect agreement with the results obtained for the
full pathway.
The use of a creation reaction for $Bar1active@extra$ is motivated
by the fact that in the full pathway this species grows linearly in
the time interval considered: by simulating the full pathway model,
we observe that at time $t=30$ it reaches the value $50$. Therefore,
its behaviour can be approximated well by a creation reaction with
rate $50/30=1.66$.

\begin{figure}[hbtp]
\centering
\includegraphics[scale=0.35]{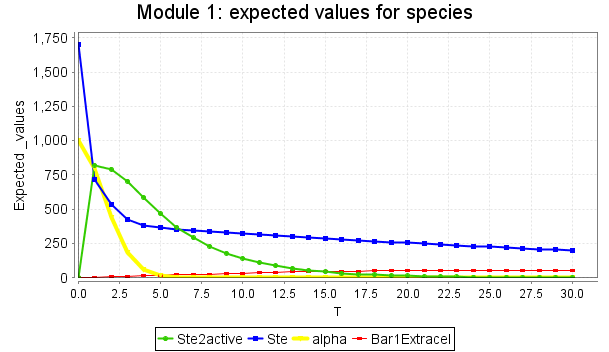}
\caption{\label{fig:module1_5}  Module 1: Expected values for the
amounts of the involved species computed in PRISM using
instantaneous reward properties.}
\end{figure}

In order to study some further properties of the module, we consider
the corresponding PRISM model. We use the simulation results to
derive the maximum amounts for the species, and we define the step
sizes $h_1=10$ for $Bar1active@extra$ and $h_2=50$ for all the other
species. The maximum numbers of levels, derived from the species'
maximum amounts and the step sizes, are $35$ for $Ste2@mem$, $20$
for $\alpha@extra$, $17$ for $Ste2active@mem$, and $5$ for
$Bar1active@extra$. Figure~\ref{fig:module1_5} reports the computed
expected values for all the species in order to validate our PRISM
model against the anticipated behaviour. The results are in
agreement with the simulation results reported in
Figure~\ref{fig:module1_ssa:full} and the results in the
literature~\cite{Kofahl04}. Some small discrepancies are due to the
choice of the step size. Indeed a finer granularity gives better
results, but the chosen one is enough to obtain some information
about the module.

\begin{figure}[hbtp]
\centering \subfigure[Prob $Ste2active@mem > 0$, $Ste2active@mem
> 16$]{\includegraphics[scale=0.33]{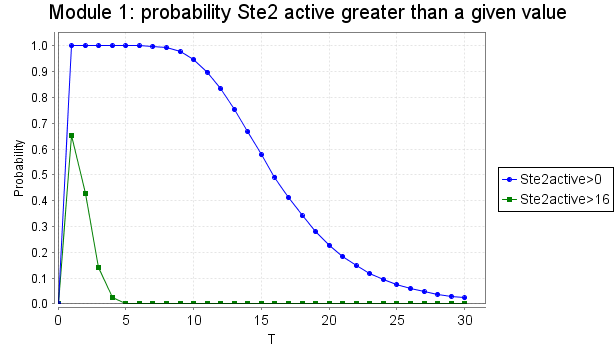}\label{fig:module1_7:a}}\hfill
\subfigure[Effect of $k_0$ on Prob $Ste2active@mem >
0$]{\includegraphics[scale=0.33]{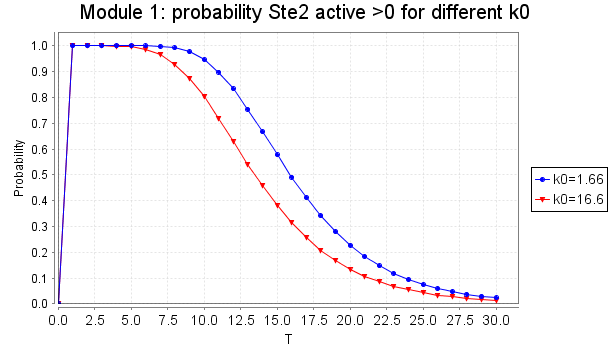}\label{fig:module1_7:b}}
\caption{\label{fig:module1_7}  Module 1: (a) comparison of the
probability of the amount of $Ste2active@mem$ being greater than
level zero and greater than level 16; (b) comparison of the
probability of the amount of $Ste2active@mem$ being greater than
zero for different values of $k_0$ (creation rate of
$Bar1active@extra$).}
\end{figure}

\begin{figure}[hbtp]
\centering
\subfigure[$Ste2active@mem/Ste2\_total$]{\includegraphics[scale=0.33]{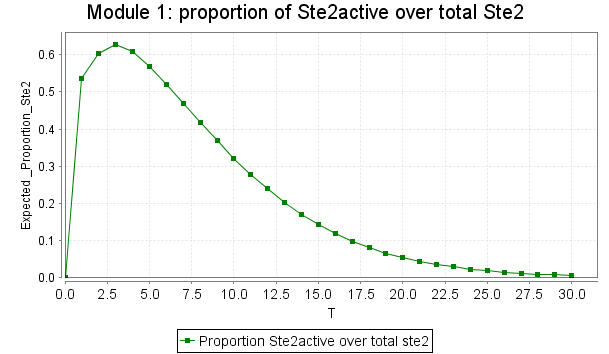}\label{fig:module1_8:a}}
\hfill \subfigure[Reaction
occurrences]{\includegraphics[scale=0.33]{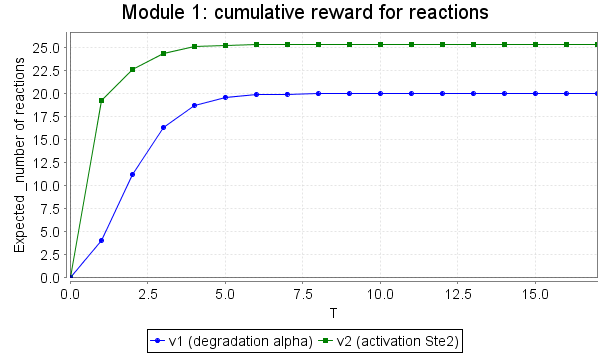}\label{fig:module1_8:b}}
\caption{\label{fig:module1_8} Module 1: (a) ratio of amount of
$Ste2active@mem$ to total amount of Ste2 ($Ste2active@mem +
Ste2@mem$); (b) expected number of reaction occurrences.}
\end{figure}

Some properties verified using PRISM are reported in
Figures~\ref{fig:module1_7} and~\ref{fig:module1_8}. We verify the
probability of $Ste2active@mem$ being greater than level zero and
greater than 16 (with level 17 corresponding to the maximum
concentration for $Ste2active@mem$) (Figure~\ref{fig:module1_7:a}),
and the probability of $Ste2active@mem$ being greater than zero for
different values of the creation rate $k_0$
(Figure~\ref{fig:module1_7:b}). Concerning the first two
probabilities (Figure~\ref{fig:module1_7:a}), in the former case the
probability rapidly grows to one and then decreases quite slowly to
zero, whereas in the latter case we soon have a peak and then the
probability rapidly decreases to zero. Indeed the activation of
$Ste2active@mem$ happens at the beginning of the pathway and then
$Ste2active@mem$ is degraded and not created anymore. Concerning the
effect on $Ste2active@mem$ of changing $k_0$
(Figure~\ref{fig:module1_7:b}), the first part is the same for all
the assumptions, but then a larger creation rate leads to a faster
decrease of the probability. The last properties we analyse regard
the ratio of $Ste2active@mem$ to $Ste2\_total = Ste2active@mem +
Ste2@mem$ (Figure~\ref{fig:module1_8:a}) and the expected number of
reactions that have been fired within a given time
(Figure~\ref{fig:module1_8:b}).

Most of the properties reported above for module 1 are valid for the
full model too and so they can be used to derive global properties for
the pathway. Indeed the results presented for this module
are in agreement with the expected behaviour of the species in the
module when considering the full pathway. This is not true when
experimenting with parameter values in the module. Indeed a change in
the parameters can lead to a variation in the amount of some species
in other modules (such as $Bar1active@extra$) and this can have an
effect on the modules themselves.

\subsection{Module 7: gene alteration}
\label{sec:module7}
\begin{figure}[hbtp]
\centering
\includegraphics[width=0.4\textwidth]{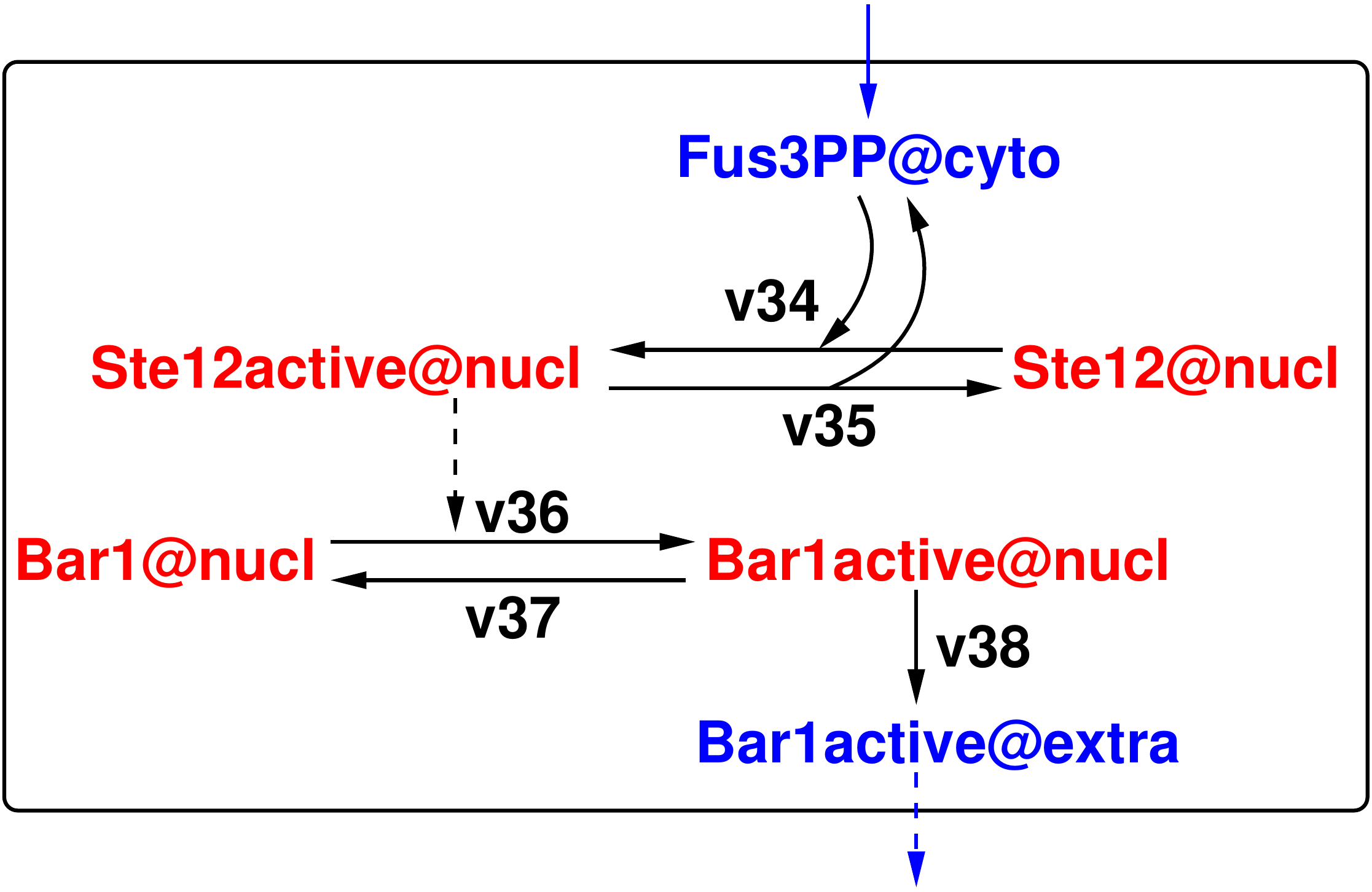}
\caption{\label{fig:module7} Description of module 7.}
\end{figure}

The second module considered in this paper is the one describing
gene alteration (Figure~\ref{fig:module7}).

This module is slightly more complex than the previous one, because
the input $Fus3PP@cyto$ is involved in several modules: in addition to
being both a reactant and a product here, $Fus3PP@cyto$ is both
reactant and product in module 5 and has a regulatory role in modules
2-3 and 6. However, the value of $Fus3PP@cyto$ is unaffected by those
modules and therefore its regulatory role in other modules is not an
issue.  The other species of this module are $Bar1active@extra$ which,
as we saw in the previous section, has a regulatory effect in module
1, and four local species: nucleic Ste12 in inactive ($Ste12@nucl$)
and active ($Ste12active@nucl$) forms, and nucleic Bar1 in inactive
($Bar1@nucl$) and active ($Bar1active@nucl$) forms.

The species definitions for the species of the module are the
following
\begin{center}
\begin{tabular}{lcl}
$Fus3PP@cyto$ & $\rmdef$ & $v_{46} \,\activator \,Fus3PP@cyto \; +  \; v_{28} \,\product \,Fus3PP@cyto \; +  \; v_{33} \,\reactant \,Fus3PP@cyto$ \\
              &          & $v_{39} \,\activator \,Fus3PP@cyto \; +  \; v_{35} \,\product \,Fus3PP@cyto \; +  \; v_{34} \,\reactant \,Fus3PP@cyto$ \\
$Ste12@nucl$ & $\rmdef$ & $v_{35} \,\product \,Ste12@nucl \; + \; v_{34} \,\reactant\, Ste12@nucl$ \\
$Ste12active@nucl$ & $\rmdef$ & $v_{35} \,\reactant \,Ste12active@nucl \; + \; v_{34} \, \product\, Ste12active@nucl \; +  \; v_{36} \, \activator \,Ste12active@nucl$ \\
$Bar1@nucl$ & $\rmdef$ & $v_{36} \, \reactant \,Bar1@nucl \; +  \; v_{37} \, \product \,Bar1@nucl$ \\
$Bar1active@nucl$ & $\rmdef$ & $v_{38} \, \reactant \,Bar1active@nucl \; + \; v_{37} \, \reactant \,Bar1active@nucl \; + \; v_{36} \, \product \,Bar1active@nucl$ \\
$Bar1active@extra$ & $\rmdef$ & $v_{38} \, \product \,Bar1active@extra \; + \; v_1 \, \activator \,Bar1active@extra$
\end{tabular}
\end{center}
\noindent and the model component describing the full module 7 is the
following.
$$\begin{array}{lcl}
Module7\_local \!\! & \!\! \rmdef \!\! & \!\! Ste12@nucl[init\_Ste12\_nucl] \syncstar{*} Ste12active@nucl[init\_Ste12active\_nucl] \syncstar{*} \\
               \!\! & \!\!        \!\! & \!\! Bar1@nucl[init\_Bar1\_nucl] \syncstar{*} Bar1active@nucl[init\_Bar1active\_nucl] \\
Module7\_I/O   \!\! & \!\! \rmdef \!\! & \!\! Fus3PP@cyto[init\_Fus3PP\_cyto] \syncstar{*}Bar1active@extra[init\_Bar1active\_extra]
\end{array}$$

\begin{figure}[hbtp]
\centering \subfigure[Full
pathway]{\includegraphics[width=0.32\textwidth]{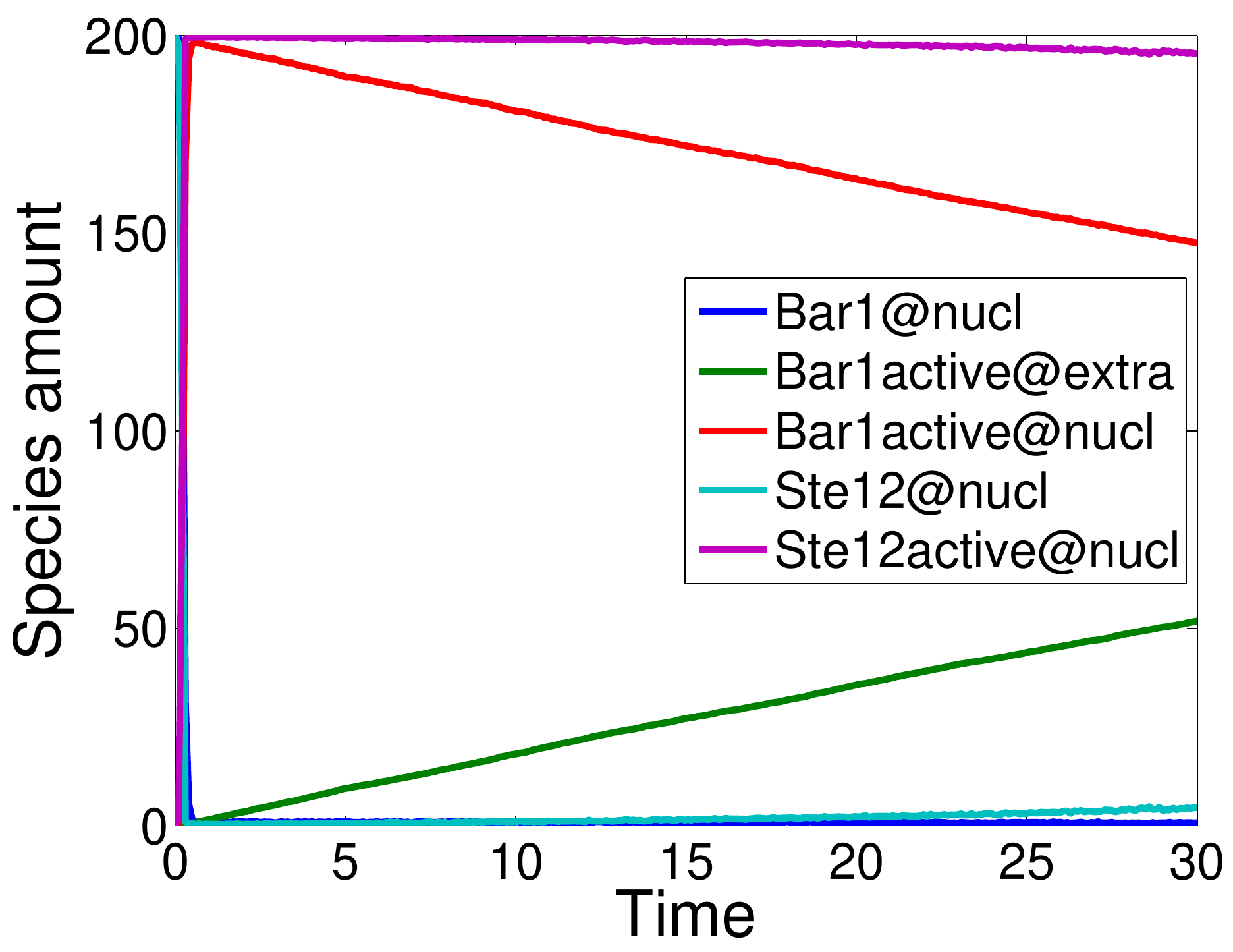}\label{fig:module7_ssa:full}}
\subfigure[$init\_Fus3PP@cyto=300$]{\includegraphics[width=0.32\textwidth]{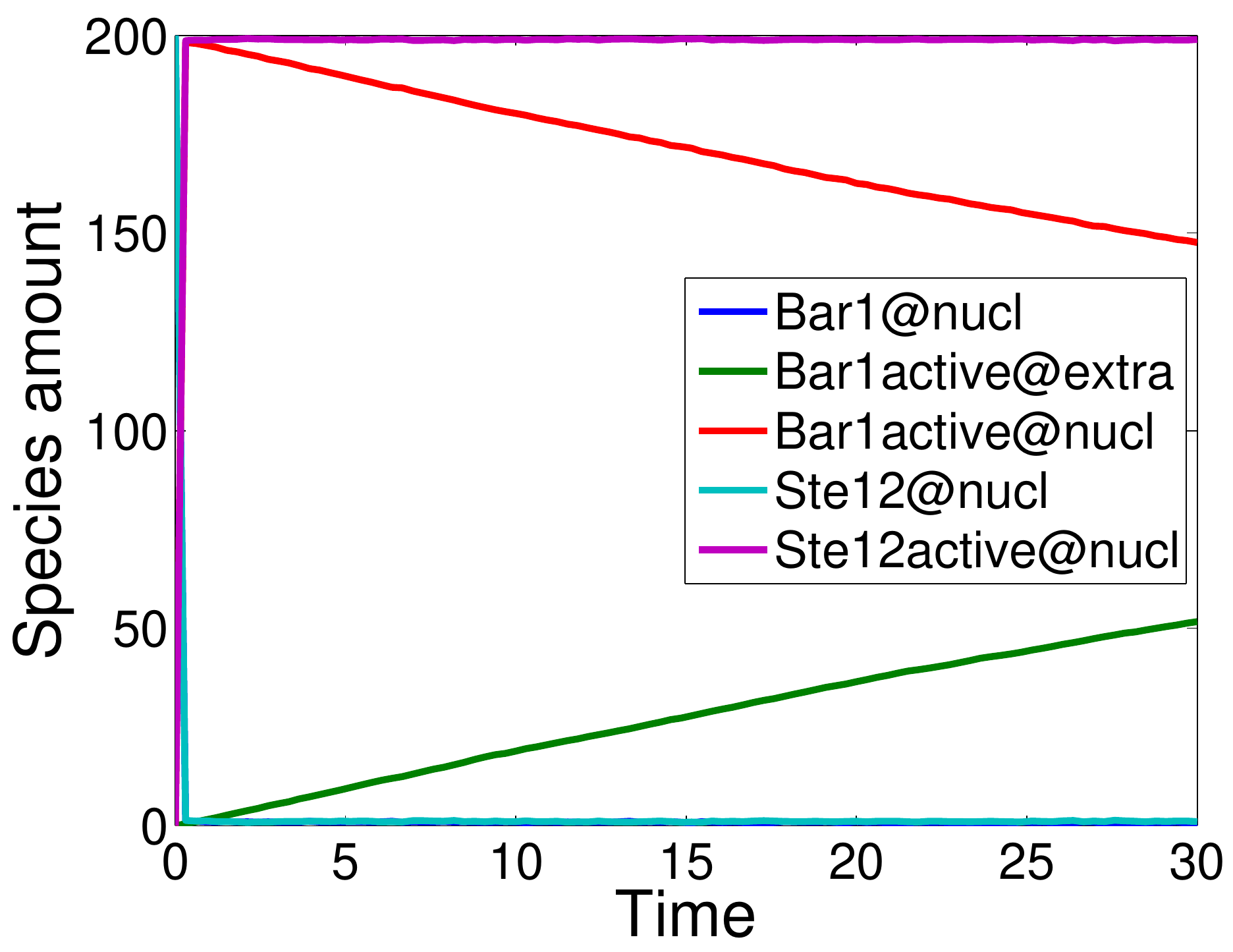}\label{fig:module7_ssa:init_300}}
\subfigure[$init\_Fus3PP@cyto=150$]{\includegraphics[width=0.32\textwidth]{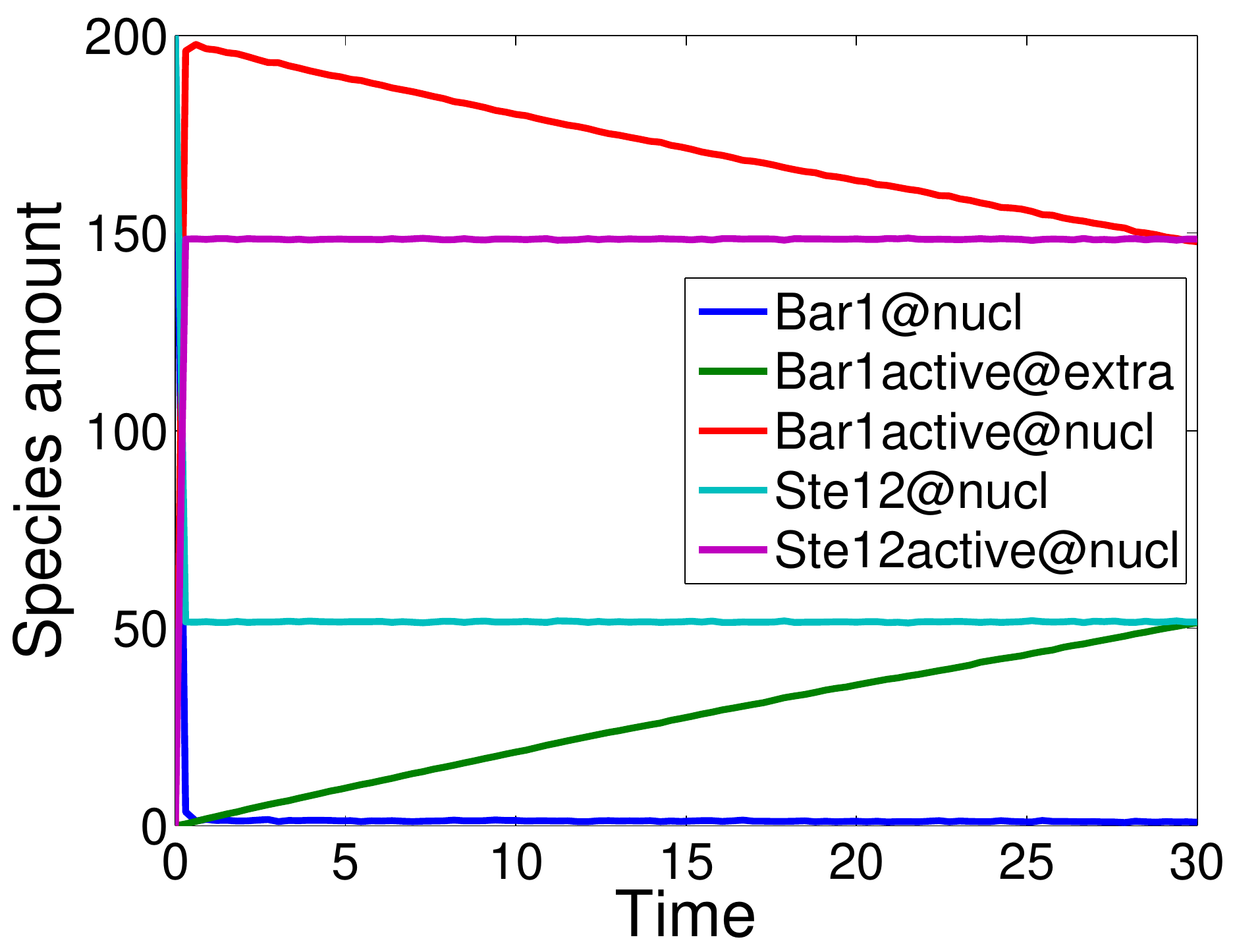}\label{fig:module7_ssa:init_150}}
\caption{\label{fig:module7_ssa} Module 7: Time-series evolution
(Gibson-Bruck method, average of 100 simulation runs) of the
involved species for different assumptions about $Fus3PP@cyto$: (a)
full pathway, (b) module 1, with initial amount of
$Fus3PP@cyto=300$, and (c) module 1, with initial amount of
$Fus3PP@cyto=150$.}
\end{figure}

As for the previous module, we have to assign a value for the
external reagent $Fus3PP@cyto$: we study the effect of its variation
on the behaviour of the module, and find an appropriate value for
which we can reproduce the results obtained on the full pathway
model. In Figure~\ref{fig:module7_ssa} we report the average
stochastic simulation results (over 100 runs) for all the species
involved in this module when the full pathway is considered
(Figure~\ref{fig:module7_ssa:full}), and for the module in isolation
(Figures~\ref{fig:module7_ssa:init_300}--\ref{fig:module7_ssa:init_150}).
Both when $Fus3PP@cyto$ is initially 300 (the maximum value it
attains in the full pathway, Figure~\ref{fig:module7_ssa:init_300})
and when it is initially 150 (half its maximum value,
Figure~\ref{fig:module7_ssa:init_150}), the behaviour of Bar1 in all
its forms is the same as the one obtained in the full pathway;
however, only using the initial value 300 the module behaves exactly
as the full pathway also in terms of the amounts $Ste12@nucl$ and
$Ste12active@nucl$.

\begin{figure}[hbtp]
\centering
\includegraphics[scale=0.34]{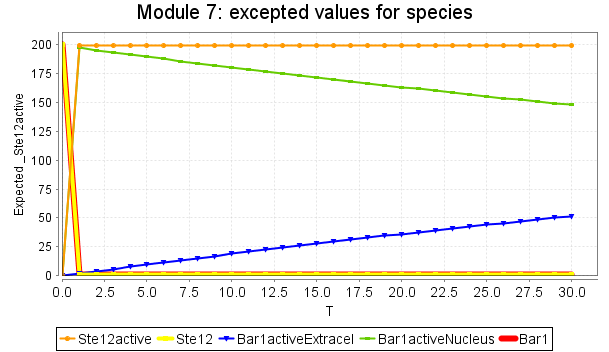}
\caption{\label{fig:module7_7} Module 7: Expected values for the
amounts of the involved species computed using PRISM.}
\end{figure}

\begin{figure}[hbtp]
\centering \subfigure[Prob $Ste12active@nucl >
0$]{\includegraphics[scale=0.35]{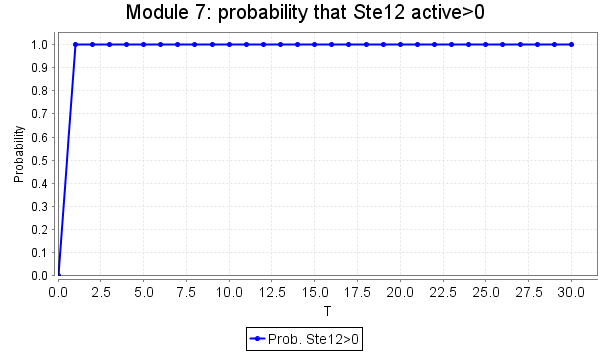}\label{fig:module7_8:a}}
\hfill \subfigure[$Ste12active@nucl / Ste12\_total$]{\includegraphics[scale=0.35]{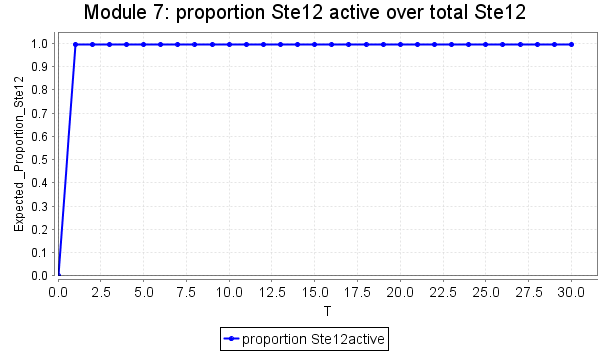}\label{fig:module7_8:b}}
\caption{\label{fig:module7_8} Module 7: (a) probability of
$Ste12active@nucl$ being greater than zero; (b) ratio of amount of
$Ste12active@nucl$ to total amount of Ste12 ($Ste12@nucl +
Ste12active@nucl$).}
\end{figure}

\begin{figure}[hbtp]
\centering
\includegraphics[scale=0.35]{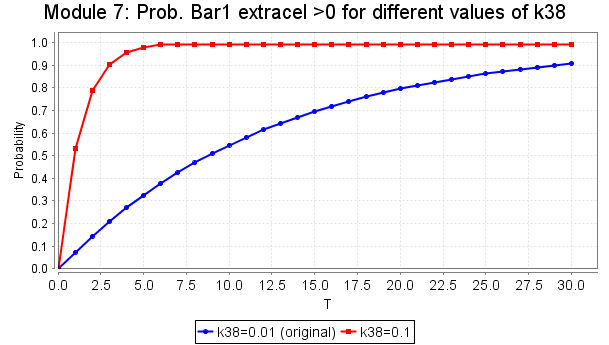}
\caption{\label{fig:module7_9} Module 7: probability of
$Ste12active@nucl$ being greater than zero for different values of
the transport rate $k_{38}$. The original value ($k_{38}=0.01$, blue
line) and $k_{38}=0.1$ (red line) are considered.}
\end{figure}

We again use PRISM to analyse the CTMC with levels associated with
the module in order to verify the satisfaction of a few desired
properties. Given the maximum amount derived from simulation,  we
define the step size $h=25$. The resulting maximum number of levels
is $12$ for $Fus3PP@cyto$ and $8$ for the other species.
Figures~\ref{fig:module7_7}, \ref{fig:module7_8},
and~\ref{fig:module7_9} report some results obtained from the
analysis in PRISM. In Figure~\ref{fig:module7_7} the expected values
for the amounts of the species are shown; these are in agreement
with the simulation results. Figure~\ref{fig:module7_8} reports the
probability of $Ste12active@nucl$ being greater than level zero
(Figure~\ref{fig:module7_8:a}), and the ratio of $Ste12active@nucl$
to $Ste12\_total = Ste12@nucl + Ste12active@nucl$
(Figure~\ref{fig:module7_8:b}). The probability of
$Ste12active@nucl$ being greater than zero increases rapidly from
zero to one and the ratio has a similar behaviour. Indeed, Ste12 is
converted rapidly from the inactive state to the active state and
the inactivation reaction is very slow compared with the activation
one. Figure~\ref{fig:module7_9} describes the probability of the
expected value of $Bar1active@extra$ being greater than zero under
different assumptions about the transport rate. As expected,
increasing the transport rate constant makes the probability grow
more rapidly to the value one. These properties, verified locally
for the module, are valid also for the full system.

\section{Conclusions}
\label{sec:conclusions}

Compositionality is a key feature of process algebras and can offer
several advantages in terms of both model construction and model
analysis, especially in the context of biological modelling. Indeed
biological systems are generally large and complex and inherently
characterised by a modular structure. Whereas there are numerous
explorations of compositional modelling using process algebras, the
application of compositionality for system analysis has not been
widely investigated.

Here we presented a preliminary study of
decomposed model analysis for biochemical systems specified in the
Bio-PEPA process algebra. We focused on the pheromone pathway in
yeast, characterised by a modular structure reflecting biological
functionalities, and we applied a modular approach for the analysis.
The approach proposed was illustrated for a single case study, but
the potential advantages, especially with respect to the increased
feasibility and efficiency of the analysis techniques, are evident.
In particular, the modular analysis allowed us to benefit from
analysis techniques, such as model-checking, whose application to
the analysis of the full system would be unfeasible due to the
state-space explosion.

This work highlighted that the definition of a general approach for
decomposed model analysis can be very useful. However this task is
challenging and several issues must be addressed. The most important
of these is the identification of suitable modules. We have shown
that modules do not need to be independent. Here we assumed that the
modules can be defined based on pre-existing biological knowledge of
the system, but it is necessary to define a general technique. 
Our examples suggest that the fine-grained modularity of the
reagent-centric modelling style supported by Bio-PEPA
\cite{CalderHill09}, offers sufficient flexibility to support
decomposition according to a wide-range of criteria.
Secondly, modules must be correctly parameterised for them to be
analysed in isolation. Here the correctness of the
modularisation and the parameterisation is performed by visual
inspection of the simulation traces obtained for the individual and
the complete model; the application for this purpose of classical
techniques for model distance and of formal relations
(e.g.~congruences) offered by process algebra is potentially interesting.
Finally, the class of system properties which can be verified
locally and are also guaranteed to hold for the full system must be
characterised. These are the main issues we are planning to
investigate in future work.

Our emphasis has been on biological decomposition and which
gives rise to a rather ad hoc mathematical decomposition. Nevertheless
we have shown that useful modular analysis can be carried out. Another
area for future work would be to start from the underlying
mathematical model, i.e.\ the CTMC, and apply established mathematical
decomposition techniques and investigate the implication for
decomposition at the biological level.

\bibliographystyle{plain} 
\bibliography{biblio}
\end{document}